\documentclass[conference]{IEEEtran}
\IEEEoverridecommandlockouts
\usepackage{cite}
\usepackage{amsmath,amssymb,amsfonts}
\usepackage{algorithmic}
\usepackage{graphicx}
\usepackage{comment}
\usepackage{subfigure}
\usepackage{textcomp}
\usepackage{xcolor}
\def\BibTeX{{\rm B\kern-.05em{\sc i\kern-.025em b}\kern-.08em
    T\kern-.1667em\lower.7ex\hbox{E}\kern-.125emX}}
\begin{document}

\title{Crafting Realistic Virtual Humans: Unveiling Perspectives on Human Perception, Crowds, and Embodied Conversational Agents}

\author{\IEEEauthorblockN{Rubens Montanha, Victor Araujo, Paulo Knob, Greice Pinho, \\
Gabriel Fonseca, Vitor Peres, Soraia Raupp Musse}
\IEEEauthorblockA{\textit{Graduate Program in Computer Science} \\
\textit{Pontifical Catholic University of Rio Grande do Sul}\\
Porto Alegre, RS, Brazil \\
Contact author: soraia.musse@pucrs.br}}

\maketitle

\begin{abstract}
Virtual Humans (VHs) were first developed more than 50 years ago and have undergone significant advancements since then. In the past, creating and animating VHs was a complex task. However, contemporary commercial and freely available technology now empowers users, programmers, and designers to create and animate VHs with relative ease. These technologies have even reached a point where they can replicate the authentic characteristics and behaviors of real actors, resulting in VHs that are visually convincing and behaviorally lifelike. This paper explores three closely related research areas in the context of virtual humans and discusses the far-reaching implications of highly realistic characters within these domains.

\end{abstract}


\section{Introduction}
\label{sec:introduction}

During the last few years, Computer Graphics (CG) have advanced in different perspectives of modeling and animation of Virtual Humans (VHs).~\footnote{Draft version made for arXiv: https://arxiv.org/}
From body scanning and use of body silhouettes \cite{magnenatthalmann2003automatic, higgins2022sympathy} to the use of highest fidelity realistic character using MetaHumans\footnote{https://metahuman.unrealengine.com/} \cite{higgins2021ascending}, it is possible to create more realistic virtual humans using the new technologies. Having more realistic characters, we can produce complicated crowd scenarios~\cite{musse2021history}, more realistic conversational agents with body animation and facial expressions~\cite{ochs2017user, sonlu2021conversational, knob2021arthur}
 and investigate aspects regarding the human perception about virtual humans~\cite{araujo2022perceived}. 

The presence of realistic virtual humans (VHs) can significantly enhance the anthropomorphic experience of users~\cite{chaminade2007anthropomorphism}, influencing their emotional responses and improving the user experience. Users naturally seek a human-like representation in VHs, encompassing various traits~\cite{garza2019emotional, zibrek2020effect}  
such as movement, emotions, and gender, which helps foster a stronger sense of connection and comfort~\cite{mori2012uncanny, 
araujo2022perceived} with these virtual beings. 
The more realistic and lifelike virtual humans (VHs) and their behaviors become, the more plausible and coherent the applications involving them become. One example is the crowd simulation area. From the pioneer works~\cite{reynolds1987flocks,musse1997model} to the most recent ones~\cite{silva2022webcrowds, chen2023agent}, many techniques have allowed creating and simulating human crowds with emotion \cite{xu2020emotion}, crowds affected by fluid forces \cite{schaffer2020towards}, abstractions of individual agents in a crowd to multi-level simulation \cite{da2019bioclouds, silva2020lodus}, a crowd estimated using machine learning methods \cite{testa2019crowdest}, among others.


Additionally, in this paper, we delve into the topic of Embodied Conversational Agents (ECA) and their application as realistic virtual humans. A considerable amount of research in this area is focused on health~\cite{spitale2020multicriteria, das2019generation} and skill training~\cite{chetty2019embodied, ayedoun2019adding}. Typically, an ECA is developed with a specific purpose in mind, such as conducting clinical interviews or facilitating skill training sessions. When considering the appearance and behavior of virtual humans (VH), verbal and non-verbal aspects are crucial. These aspects encompass eye movements~\cite{lee2002eyes}, empathetic behavior~\cite{yalccin2020empathy}, personality~\cite{sajjadi2019personality}, memory~\cite{martinez2020multiparty}, among others. Given that ECAs are designed to be user-friendly interfaces that incorporate empathy and foster excellent user interaction, pursuing more realistic VHs significantly impacts this field.



This paper is organized as follows. In Section~\ref{sec:visual_perception}, we discuss various aspects of visual perception, addressing topics like the Uncanny Valley and biases in virtual human perception. Section~\ref{sec:app} focuses on the application areas of crowd simulation and embodied conversational agents. Both Sections~\ref{sec:visual_perception} and~\ref{sec:app} thoroughly explore the fundamentals and relevant literature in their respective domains. Finally, in Section~\ref{sec:final_considerations}, we present our concluding remarks, which encompass new emerging trends and potential future directions for further research and development.

\section{Visual Perception of Virtual Humans}
\label{sec:visual_perception}

Perception, stemming from the Latin word meaning "to apprehend" or "to understand," refers to the brain's process of interpreting sensory information and giving it meaning and context~\cite{andreotti2021perception}. 
In the context of CG, perception plays a crucial role in how virtual environments and characters are created and perceived by users. CG techniques often rely on understanding how the human visual system interprets visual stimuli to achieve more realistic and convincing results~\cite{zell2019perception}. 
By considering human perception, researchers and artists can design virtual humans and environments that align with how real-world stimuli are interpreted by the human brain. 
More specifically, researchers can improve rendering algorithms, animation methods, and modeling techniques to create virtual humans that appear more natural and evoke emotional connections with users~\cite{katsyri2015review, zibrek2020effect}. Perception remains highly relevant when discussing virtual humans' evolution and CG' advancement. Understanding how humans perceive and interpret visual information is crucial for enhancing the realism, believability, and overall quality of virtual characters and environments. As CG continues to progress, the consideration of human perception will remain an essential aspect in creating more immersive and compelling virtual experiences.

\subsection{Uncanny Valley Concepts}
\label{sec:uv_concepts}

\begin{figure*}[!htb]
 \centering
 \subfigure
 {\includegraphics[width=0.47\textwidth]{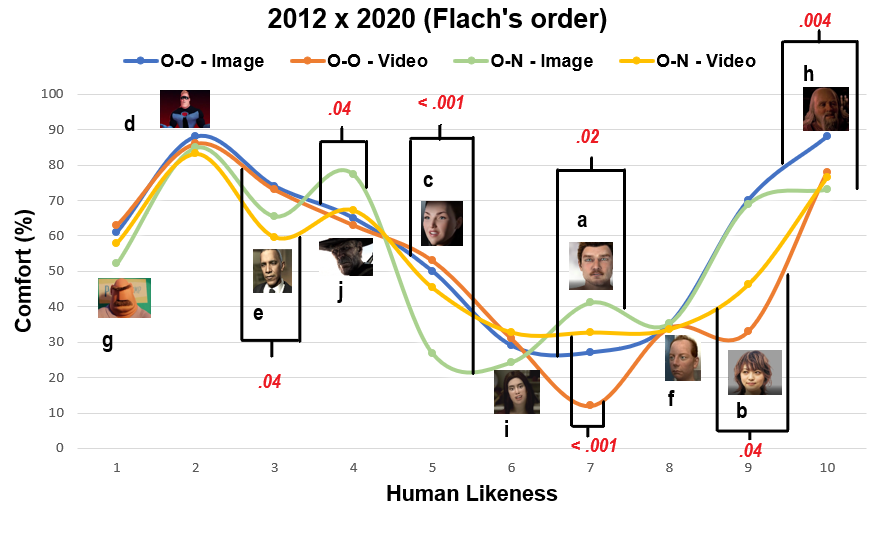}}
 \subfigure
 {\includegraphics[width=0.47\textwidth]{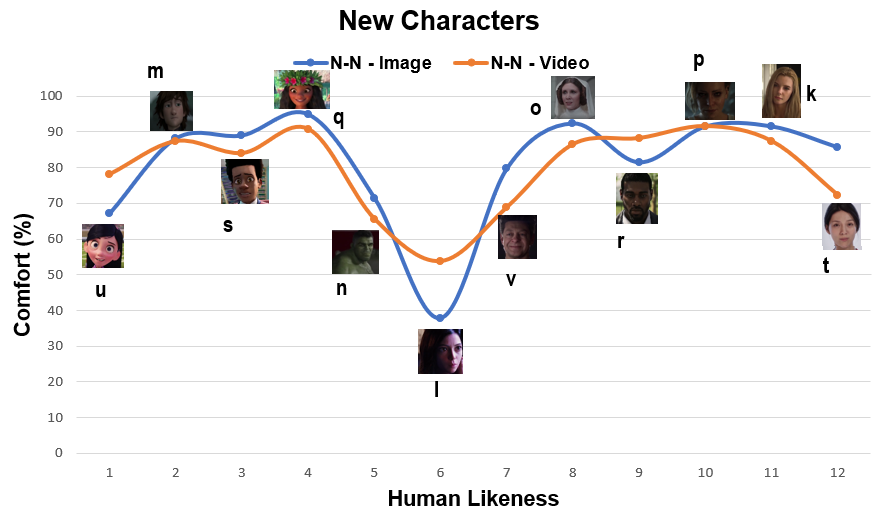}}
  \caption{All the characters used in the work of Flach et al.~\cite{dill2012evaluation} (left), and the perceptual comfort obtained with more recent characters (right). Both blue and orange lines on the left represent the percentages of the perceived comfort of each character in the image and video, as perceived in 2012. However, the green and yellow lines represent the same VHs evaluated in 2020. In addition, in (b), we can see the results regarding recent characters perceived in 2020~\cite{araujo2022perceived}}.
  \label{fig:ieee}
\end{figure*}

Realistic VHs from movies and games can cause strangeness and involuntary feelings in viewers, an effect known as the Uncanny Valley (UV). 
According to Mori~\cite{mori2012uncanny}, who proposed the UV, robots that appear too similar to real humans can fall into the UV, where sometimes a high degree of human realism evokes an eerie feeling in the viewer.  
The sense of discomfort perceived in certain virtual characters, as discussed in the UV theory, can be a critical factor in our perceptual and cognitive discrimination. Assessing the perceived quality of the content of images and videos is essential in processing this data in various applications, such as films, games, and platforms that use images to communicate relevant information~\cite{shahid2014no}. The area of visual perception is highly complex, influenced by many factors, not fully understood, and challenging to model and measure. 
For Tinwell et al.~\cite{tinwell2011facial}, 
the phenomenon UV means that virtual characters too similar to humans can evoke an adverse reaction from the observer, as they have an appearance and behavior different from what would be considered a typical pattern in humans. Other studies in 
show characteristics of VHs that are already considered more strange to humans when evaluated, such as actions perceived as unnatural, rigid, or abrupt movements, as shown in the study by Bailenson et al.~\cite{bailenson2005independent}; lack of human similarity in the speech and facial expression of a character, in the studies by Tinwell et al.~\cite{tinwell2011facial}; lip synchronization error that can be expressed before lip movement or vice versa, according to the studies by Gouskos et al.~\cite{gouskos2006depths}.
%
In a work by Araujo et al.~\cite{araujo2022perceived}, the authors conducted a perceptual study involving the UV theory and CG characters from different media, such as movies, games, etc. The authors recreated the work of Dill et al.~\cite{dill2012evaluation} and compared the average comfort perceived by people in 2012 about characters from 2012 and the average comfort perceived by new people in 2021 for the same characters older than 2012. The results showed indications that the average comfort about the new characters was higher (statistically) than that of the old characters. 
Figure~\ref{fig:ieee} illustrates some results of users' perceived comfort when watching VH humans' images and videos~\cite{araujo2022perceived}. On the left, we see the comparison between the perceived comfort of people in 2012 and 2020, while on the right, we can see people's comfort levels regarding more recent VHs. It is easy to notice that people perceive more comfort in recent VHs than in the old ones. It is explained by the advancement of technology and may indicate that more realistic characters will positively impact the users' experience.

\subsection{Computational Uncanny Valley}
\label{sec:computational_uv}

As mentioned in the last section, the area of visual perception is highly complex 
and challenging to model and measure~\cite{beghdadi2013survey}. For these reasons, subjective assessments are still very used, in which a group of human viewers evaluates the images/videos~\cite{Theodoridis2014} qualitatively. 
However, some applications may require objective assessment when subjective analysis is impossible. 
Dal Molin et al.~\cite{dal2021can} 
present a study on image analysis 
which aims to assess the quantitative comfort of animated characters' faces. It was not the first time image quality was quantitatively measured using computational methods. However, it was the first time the quality of the image to be assessed was the feeling of strangeness giving a VH face. 
Starting from the conceptualization of Tumblin and Ferwerda~\cite{tumblin2001applied} that perception is a set of processes that actively build mental representations of the world, Dal Molin et al.~\cite{dal2021can} investigate the visual features as a way to detect strangeness perceived by human beings when observing CG faces. One of these features is the saliency mapping that highlights the most relevant face lines in the characters' faces. Two other resources for extracting information are Hu Moments and Hog. The first one calculates the weight of the pixel intensity, while the second one describes the distribution of intensity gradients and edge directions. We also use the impact of the PCA feature as a dimensionality reducer to use only relevant variables. The authors use a Support Vector Machine (SVM) for binary classification of the character's comfort and compare it with the estimation of strangeness perceived by subjects, where class 0 indicates that the character does not cause strangeness and class 1 is the opposite. The best accuracy (80\%) was obtained using Hu moments, without saliency, Polynomial kernel, and  PCA for dimensionality reduction. 

In a previous work~\cite{dal2022estimating}, we proposed a model for estimating the comfort value (continuous) a specific CG face should cause in humans' perception. As dealt by Liu et al~\cite{liu2014no}, 
we use spatial and spectral entropy to estimate image quality. 
We introduced $CCS$ as the Computed Comfort Score and tested the same faces as in~\cite{dal2021can} 
to check for evidence of accuracy, constantly confronting results with subjects' opinions. We obtained an accuracy of 80\% when using $CCS$ to classify the characters in a binary classification (comfort/discomfort) and an MAE of 23.59\% when comparing the $CCS$ values with subjective data. These results indicate that estimating people's subjective perception through computational methods seems possible. Indeed, one question here is if the increasing realism of Virtual Humans impacts our results. 
This is a future work we intend to pursue.


\subsection{Bias in Virtual Humans Perception}
\label{sec:bias_vh}

Laue~\cite{laue2017familiar} states that we, as a species, tend to anthropomorphize (through bias indicators~\cite{kteily2022dehumanization}) and sympathize, even with the simplest and weirdest technologies. In addition, from the designer to the viewer, people tend to attribute human characteristics, such as gender, skin color, race, etc, to these technologies. 
There are several perceptive studies that address assessments regarding gender bias and virtual humans~\cite{araujo2022towards, mcdonnell2007virtual, zibrek2020effect, araujo2021analysis}. 
In the work of Araujo et al.~\cite{araujo2021analysis}, the authors conducted a perceptive study to compare UV theory from a gender perspective. 
The results showed that women tend to feel more comfortable with realistic female CG characters than men. In addition, the study showed indications that women and men could categorize the UV chart in different ways depending on the gender of the VHs. 
Therefore, in the context of CG, some work in the literature has also discussed that. One example aims to eliminate substantial evidence of biases~\cite{kim2022countering} when incorporating gender into the design of VHs. It is crucial for professionals such as designers, programmers, and researchers to eradicate their own gender biases to deconstruct gender stereotypes~\cite{araujo2022towards}. Moreover, they should consider the diversity of the audience's gender identities and experiences, visually and behaviorally~\cite{mcdonnell2007virtual}, to create inclusive and representative VHs.

In the research conducted by Araujo et al.~\cite{araujo2022towards}, the authors aimed to measure gender bias using a virtual baby (VB) as a genderless virtual character. The study replicated Condry and Condry's \cite{condry1976sex} psychology study, which considered babies inherently genderless due to their lack of social and cultural biases. Araujo et al. presented videos of a VB interacting with various toys to three different participant groups. To attribute gender to the VB, the authors assigned names to two groups: one group received a female name, another group a male name, while the third group observed the VB without any assigned name. Participants were then asked to evaluate the possible emotions displayed by the VB during the videos and to assign a gender to the character. The findings of the study were twofold. First, the authors suggested that attributing gender to genderless virtual characters could serve as a potential solution to reduce gender stereotypes in virtual human design. Second, they observed that people's gender biases, as seen in real-life interactions, extended to virtual environments. In conclusion, Araujo et al.'s research shed light on the significance of gender attribution to virtual characters and the transfer of real-life biases into virtual settings. 
These findings raise critical awareness about the need for diversity and inclusivity in CG research and VHs design. Additionally, all these aspects are more relevant in realistic VHs, where users can be identified and feel represented. 
This consideration is essential for addressing the underrepresentation and potential bias in current virtual human design and ensuring that technology reflects the diversity of the real world.



\section{Virtual Humans Applications}
\label{sec:app}

In this section, we discuss two areas of application with VHs. Firstly, we discuss crowd simulation and some challenges to obtaining realism regarding VHs behaviors. Then, we present some characteristics and research developed in the context of ECAs (Embodied Conversational Agents).

\subsection{Crowd Simulation}
\label{sec:crowd_simulation}

Crowd simulation techniques find widespread application in representing the movement of populations within both interior and exterior environments, catering to various levels of detail. Microscopic models view each agent as an individual entity with homogeneous behaviors, exemplified by the Social Forces Model introduced by Helbing and Molnár~\cite{helbing1995social}. 
Models like ORCA by Van den Berg et al.~\cite{van2011reciprocal}, and BioCrowds by Bicho et al.~\cite{de2012simulating} offer collision avoidance mechanisms and self-organizing crowd behaviors through a competitive space partitioning approach. 
In contrast, microscopic models with heterogeneous behaviors explore individual agent variations. Pelechano et al.~\cite{pelechano2005crowd} incorporate psychological aspects and communication to generate emergent cultural behavior, while Durupinar et al.~\cite{durupinar2015psychological} differentiate passive and active crowds based on emotional and homogeneous behavior, employing personality traits and emotion models for multi-agent simulations. 

\begin{figure*}[!htb]
 \centering
 \subfigure
 {\includegraphics[width=0.27\textwidth]{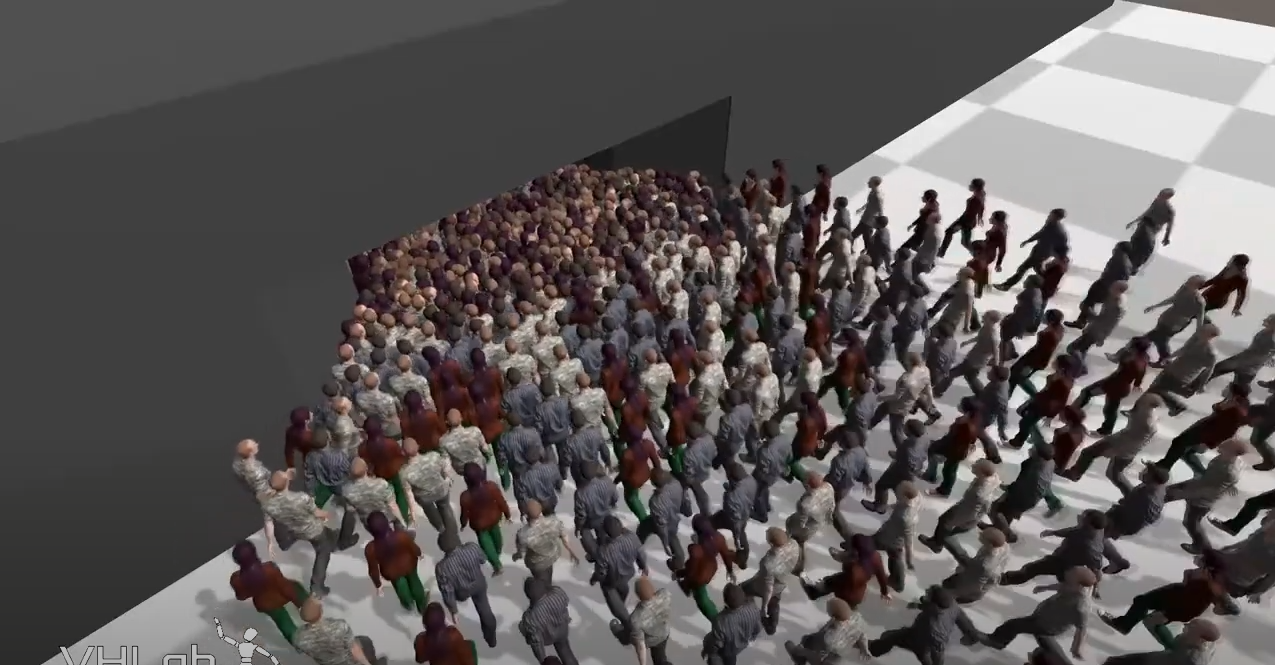}}
 \subfigure
 {\includegraphics[width=0.368\textwidth]{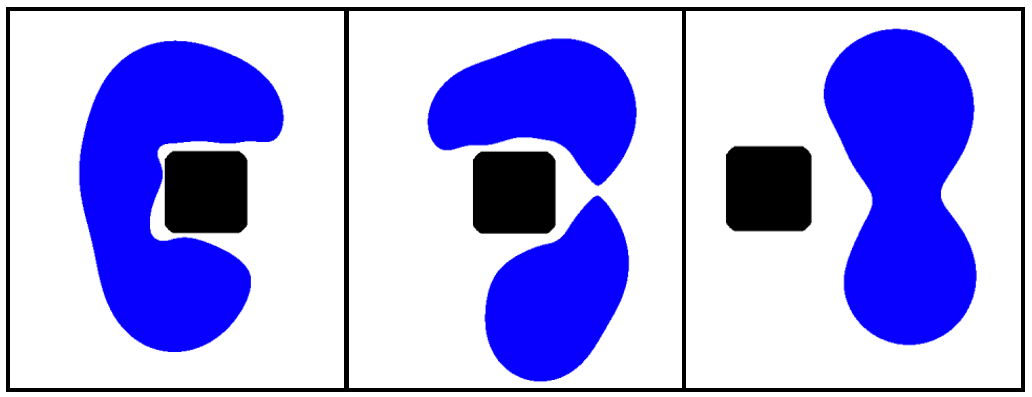}}
 \subfigure
 {\includegraphics[width=0.31\textwidth]{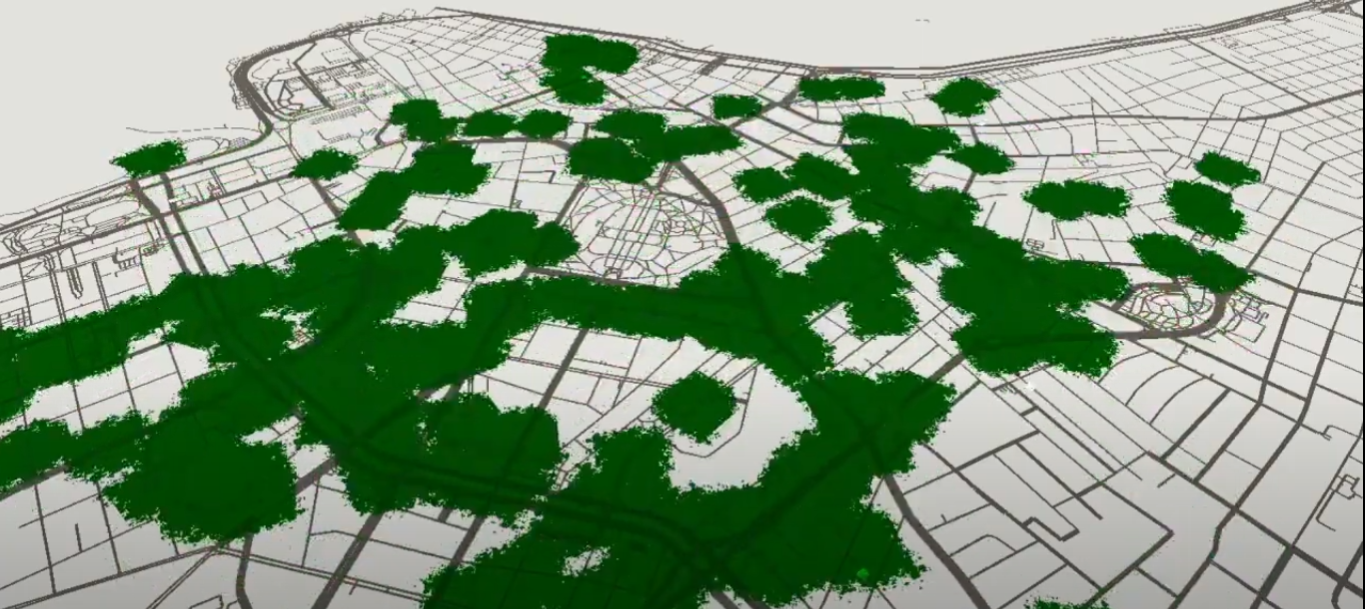}}
  \caption{On the left, we observe a bottleneck scenario where the microscopic crowd is densely organized (people/$m^2$) as they pass through the door. In the center, we present an example of a macroscopic crowd simulation, demonstrating how the crowd avoids collisions with objects by dynamically splitting and merging before and after the collision event, respectively~\cite{da2020towards}. On the right, we glimpse a city designed to serve as the environment for a macroscopic population simulation~\cite{silva2020lodus}.} 
  \label{fig:crowds}
\end{figure*}

Macroscopic crowd simulation models, on the other hand, represent agents as part of a collective influenced by a macro-level structure, offering higher granularity with a trade-off in individual characteristics. 
Antonitsch et al.\cite{da2019bioclouds} introduce BioClouds, a cloud-based approach derived from BioCrowds, where clouds compete for space based on desired densities of VHs. Hybrid models, striving for accuracy and computational efficiency, combine both microscopic and macroscopic approaches. Xiong et al.\cite{xiong2010hybrid} divide the environment into static regions simulated with ORCA and Continuum Crowds, facilitating smooth agent transitions between boundaries. Similarly, BioClouds and Legion~\cite{da2019bioclouds, da2020towards} enable seamless transitions between macroscopic and microscopic simulations, dynamically adjusting agent level of detail according to environmental complexity while facilitating the splitting and merging of legions. Figure~\ref{fig:crowds} depicts crowd simulation images. On the left, BioCrowds generates virtual humans navigating through a bottleneck, experiencing a constriction of available walking space, leading to high VH densities in the environment during a microscopic simulation. In the center, we show a blob visualization of a macroscopic Legion of individuals~\cite{da2020towards}, while on the right, a BioClouds simulation presents a cloud-based representation with no individuals~\cite{da2019bioclouds}. Microscopic crowd simulations excel in simulating numerous individuals, yet complex individual behaviors such as facial animations or small actions remain challenging due to computational constraints. Realistic crowd behavior control and the computational aspect present a significant challenge. However, the prospect of incorporating more realistic virtual humans in crowd simulations holds immense potential for applications requiring heightened realism.

\subsection{Embodied Conversational Agents}
\label{sec:eca}

%
Embodied Conversational Agents (ECAs) are virtual humans who can interact and talk with humans naturally. The main difference between an ECA and a chatbot, like ChatGPT ((Chat Generative Pre-Trained Transformer))~\cite{radford2018improving}, is that an ECA is endowed with some level of embodiment, for instance, a head, a face, and a body, so an ECA also has visual behaviors in addition to textual communication. In this case, the ECAs used technologies employed to build VHs if human characters represent them. In recent years, much research has aimed to improve the quality of the verbal and non-verbal communication abilities of Embodied Conversational Agents, both verbal and non-verbal~\cite{yalccin2020empathy,biancardi2019computational}. 
A fair amount of effort is being directed on ECAs, which can help people to have a healthier life~\cite{kramer2019developing,spitale2020multicriteria} 
for clinical interviews~\cite{philip2020trust} 
and the training of some skill~\cite{chetty2019embodied}. 

\begin{figure*}
\includegraphics[width=\textwidth]{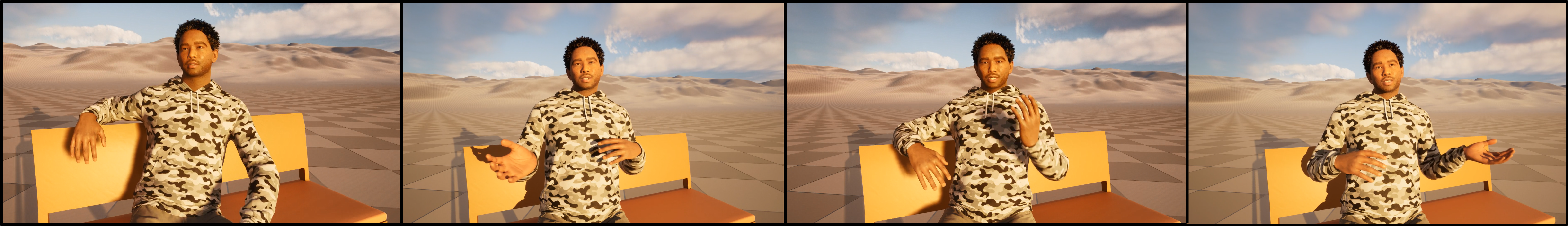}
  \caption{Illustrations of facial and body animations designed for Arthur using Unreal and MetaHuman.}
  \label{fig:teaser}
\end{figure*}

Since an ECA is endowed with a face, there is a concern about how this face (and the VH as a whole) is presented to the user. As shown in Dill et al.~\cite{dill2012evaluation}, when trying to present a character that pretends to be human-like, there is a certain eerie feeling (UV) when such virtual agent looks human, but not as close as expected from a real one, especially when animation is poorly made. Also, considering a character’s face, the authors affirm that the elements that can cause more strangeness to people are eyes and mouths. Therefore, the authors conclude that people tend to have a better reception of cartoon-like characters, or a VH, which is very close to what is expected as a human being, in terms of realism. Concerning these high-fidelity virtual humans, one of the most used tools for building such characters is MetaHuman\footnote{https://www.unrealengine.com/en-US/metahuman\label{fot:mh}}, developed by Unreal.
Knob et al.~\cite{knob2021arthur} proposed an Embodiment Conversational Agent named Arthur, endowed with a human-like memory model, which was used to improve communication. It could also recognize the person it was talking to and detect his/her expressed emotions through facial expressions. In more recent work, Knob et al.~\cite{knob2023arthur} introduced Bella, which was also endowed with empathetic behavior, and a connection between the said empathy and its memory module. 
Figure~\ref{fig:teaser} illustrates a new version of Arthur, implemented using Unreal and MetaHuman\footref{fot:mh}. 

\section{Challenges and Final Considerations}
\label{sec:final_considerations}

In recent years, the field of modeling and animation of virtual humans has witnessed remarkable advancements, enabling the creation of highly sophisticated and realistic avatars. These virtual humans are becoming increasingly indistinguishable from real individuals, thanks to advancements in computer graphics, machine learning, and motion-capture technologies. As a result, users' perception of virtual humans has evolved significantly, with users now experiencing a stronger sense of presence and emotional connection with these virtual beings. However, alongside this progress comes the challenge of ensuring ethical use and avoiding the so-called 'uncanny valley' effect, where virtual humans appear almost realistic but exhibit subtle inconsistencies that can lead to discomfort for users. Future research in this area should focus on fine-tuning virtual human models to bridge this gap and exploring novel methods to personalize virtual human interactions, allowing for more adaptive and engaging experiences.

The applications of virtual humans in crowds and embodied conversational agents are emerging as transformative areas of research and practical implementation. In crowds simulation, virtual humans are increasingly being utilized to simulate realistic and dynamic scenarios, enhancing training simulations and improving urban planning and crowd management strategies. Moreover, they play a crucial role in gaming and entertainment, creating immersive virtual worlds populated by lifelike characters. Similarly, embodied conversational agents, equipped with natural language processing and emotional understanding, have shown the way for more natural and engaging human-computer interactions. From customer service and virtual companions to educational tools, embodied conversational agents have the potential to revolutionize various domains, making interactions with technology more intuitive and emotionally meaningful. As these applications continue to evolve, future directions should focus on ensuring inclusivity and cultural sensitivity in the design of virtual humans for diverse contexts and advancing the integration of AI-driven algorithms to enable more dynamic and context-aware interactions. 

\section*{Acknowledgment}
The authors would like to thank Capes, CNPq, and FAPERGS, who financially support this research. 

\bibliographystyle{IEEEtran}
\bibliography{bib}

\begin{thebibliography}{10}
\providecommand{\url}[1]{#1}
\csname url@samestyle\endcsname
\providecommand{\newblock}{\relax}
\providecommand{\bibinfo}[2]{#2}
\providecommand{\BIBentrySTDinterwordspacing}{\spaceskip=0pt\relax}
\providecommand{\BIBentryALTinterwordstretchfactor}{4}
\providecommand{\BIBentryALTinterwordspacing}{\spaceskip=\fontdimen2\font plus
\BIBentryALTinterwordstretchfactor\fontdimen3\font minus \fontdimen4\font\relax}
\providecommand{\BIBforeignlanguage}[2]{{%
\expandafter\ifx\csname l@#1\endcsname\relax
\typeout{** WARNING: IEEEtran.bst: No hyphenation pattern has been}%
\typeout{** loaded for the language `#1'. Using the pattern for}%
\typeout{** the default language instead.}%
\else
\language=\csname l@#1\endcsname
\fi
#2}}
\providecommand{\BIBdecl}{\relax}
\BIBdecl

\bibitem{magnenatthalmann2003automatic}
N.~Magnenat-Thalmann, H.~Seo, and F.~Cordier, ``Automatic modeling of animatable virtual humans-a survey,'' in \emph{Fourth International Conference on 3-D Digital Imaging and Modeling, 2003. 3DIM 2003. Proceedings.}\hskip 1em plus 0.5em minus 0.4em\relax IEEE, 2003, pp. 2--10.

\bibitem{higgins2022sympathy}
D.~Higgins, K.~Zibrek, J.~Cabral, D.~Egan, and R.~McDonnell, ``Sympathy for the digital: Influence of synthetic voice on affinity, social presence and empathy for photorealistic virtual humans,'' \emph{Computers \& Graphics}, vol. 104, pp. 116--128, 2022.

\bibitem{higgins2021ascending}
D.~Higgins, D.~Egan, R.~Fribourg, B.~Cowan, and R.~McDonnell, ``Ascending from the valley: Can state-of-the-art photorealism avoid the uncanny?'' in \emph{ACM Symposium on Applied Perception 2021}, 2021, pp. 1--5.

\bibitem{musse2021history}
S.~R. Musse, V.~J. Cassol, and D.~Thalmann, ``A history of crowd simulation: the past, evolution, and new perspectives,'' \emph{The Visual Computer}, pp. 1--16, 2021.

\bibitem{ochs2017user}
M.~Ochs, C.~Pelachaud, and G.~Mckeown, ``A user perception--based approach to create smiling embodied conversational agents,'' \emph{ACM Transactions on Interactive Intelligent Systems (TiiS)}, vol.~7, no.~1, pp. 1--33, 2017.

\bibitem{sonlu2021conversational}
S.~Sonlu, U.~G{\"u}d{\"u}kbay, and F.~Durupinar, ``A conversational agent framework with multi-modal personality expression,'' \emph{ACM Transactions on Graphics (TOG)}, vol.~40, no.~1, pp. 1--16, 2021.

\bibitem{knob2021arthur}
P.~Knob, W.~S. Dias, N.~Kuniechick, J.~Moraes, and S.~R. Musse, ``Arthur: a new eca that uses memory to improve communication,'' in \emph{2021 IEEE 15th International Conference on Semantic Computing (ICSC)}.\hskip 1em plus 0.5em minus 0.4em\relax IEEE, 2021, pp. 163--170.

\bibitem{araujo2022perceived}
V.~Araujo, J.~Melgare, B.~M. Dalmoro, and S.~R. Musse, ``Is the perceived comfort with cg characters increasing with their novelty?'' \emph{IEEE Computer Graphics and Applications}, vol.~42, no.~1, pp. 32--46, 2021.

\bibitem{chaminade2007anthropomorphism}
T.~Chaminade, J.~Hodgins, and M.~Kawato, ``Anthropomorphism influences perception of computer-animated characters’ actions,'' \emph{Social cognitive and affective neuroscience}, vol.~2, no.~3, pp. 206--216, 2007.

\bibitem{garza2019emotional}
M.~Garza, E.~Akleman, S.~Harris, and F.~House, ``Emotional silence: Are emotive expressions of 3d animated female characters designed to fit stereotypes,'' in \emph{Women's Studies International Forum}, vol.~76.\hskip 1em plus 0.5em minus 0.4em\relax Elsevier, 2019, p. 102252.

\bibitem{zibrek2020effect}
K.~Zibrek, B.~Niay, A.-H. Olivier, L.~Hoyet, J.~Pettre, and R.~McDonnell, ``The effect of gender and attractiveness of motion on proximity in virtual reality,'' \emph{ACM Transactions on Applied Perception (TAP)}, vol.~17, no.~4, pp. 1--15, 2020.

\bibitem{mori2012uncanny}
M.~Mori, K.~F. MacDorman, and N.~Kageki, ``The uncanny valley [from the field],'' \emph{IEEE Robotics \& Automation Magazine}, vol.~19, no.~2, pp. 98--100, 2012.

\bibitem{reynolds1987flocks}
C.~W. Reynolds, ``Flocks, herds and schools: A distributed behavioral model,'' in \emph{Proceedings of the 14th annual conference on Computer graphics and interactive techniques}, 1987, pp. 25--34.

\bibitem{musse1997model}
S.~R. Musse and D.~Thalmann, ``A model of human crowd behavior: Group inter-relationship and collision detection analysis,'' in \emph{Computer Animation and Simulation’97: Proceedings of the Eurographics Workshop in Budapest, Hungary, September 2--3, 1997}.\hskip 1em plus 0.5em minus 0.4em\relax Springer, 1997, pp. 39--51.

\bibitem{silva2022webcrowds}
G.~F. Silva, P.~Knob, R.~Montanha, and S.~Musse, ``Webcrowds: An authoring tool for crowd simulation,'' in \emph{2022 21st Brazilian Symposium on Computer Games and Digital Entertainment (SBGames)}.\hskip 1em plus 0.5em minus 0.4em\relax IEEE, 2022, pp. 1--6.

\bibitem{chen2023agent}
Y.~Chen, C.~Wang, X.~Du, Y.~Shen, and B.~Hu, ``An agent-based simulation framework for developing the optimal rescue plan for older adults during the emergency evacuation,'' \emph{Simulation Modelling Practice and Theory}, p. 102797, 2023.

\bibitem{xu2020emotion}
M.~Xu, C.~Li, P.~Lv, W.~Chen, Z.~Deng, B.~Zhou, and D.~Manocha, ``Emotion-based crowd simulation model based on physical strength consumption for emergency scenarios,'' \emph{IEEE Transactions on Intelligent Transportation Systems}, vol.~22, no.~11, pp. 6977--6991, 2020.

\bibitem{schaffer2020towards}
D.~Schaffer, A.~Antonitsch, A.~Neto, and S.~Musse, ``Towards animating virtual humans in flooded environments,'' in \emph{Proceedings of the 13th ACM SIGGRAPH Conference on Motion, Interaction and Games}, 2020, pp. 1--10.

\bibitem{da2019bioclouds}
A.~Da~Silva~Antonitsch, D.~H.~M. Schaffer, G.~W. Rockenbach, P.~Knob, and S.~R. Musse, ``Bioclouds: A multi-level model to simulate and visualize large crowds,'' in \emph{Advances in Computer Graphics: 36th Computer Graphics International Conference, CGI 2019, Calgary, AB, Canada, June 17--20, 2019, Proceedings 36}.\hskip 1em plus 0.5em minus 0.4em\relax Springer, 2019, pp. 15--27.

\bibitem{silva2020lodus}
G.~F. Silva, V.~Cassol, A.~B.~F. Neto, A.~Antonitsch, D.~Schaffer, S.~R. Musse, and R.~de~Marsillac~Linn, ``Lodus: A multi-level framework for simulating environment and population-a contagion experiment on a pandemic world,'' in \emph{2020 IEEE International Smart Cities Conference (ISC2)}.\hskip 1em plus 0.5em minus 0.4em\relax IEEE, 2020, pp. 1--8.

\bibitem{testa2019crowdest}
E.~Testa, R.~C. Barros, and S.~R. Musse, ``Crowdest: a method for estimating (and not simulating) crowd evacuation parameters in generic environments,'' \emph{The Visual Computer}, vol.~35, pp. 1119--1130, 2019.

\bibitem{spitale2020multicriteria}
M.~Spitale, F.~Catania, P.~Crovari, and F.~Garzotto, ``Multicriteria decision analysis and conversational agents for children with autism,'' in \emph{Proceedings of the 53rd Hawaii International Conference on System Sciences}, 2020.

\bibitem{das2019generation}
K.~S.~J. Das, T.~Beinema, H.~Op~Den~Akker, and H.~Hermens, ``Generation of multi-party dialogues among embodied conversational agents to promote active living and healthy diet for subjects suffering from type 2 diabetes,'' in \emph{5th International Conference on Information and Communication Technologies for Ageing Well and e-Health, ICT4AWE 2019}.\hskip 1em plus 0.5em minus 0.4em\relax SCITEPRESS, 2019, pp. 297--304.

\bibitem{chetty2019embodied}
G.~Chetty and M.~White, ``Embodied conversational agents and interactive virtual humans for training simulators,'' in \emph{Proc. The 15th International Conference on Auditory-Visual Speech Processing}, 2019, pp. 73--77.

\bibitem{ayedoun2019adding}
E.~Ayedoun, Y.~Hayashi, and K.~Seta, ``Adding communicative and affective strategies to an embodied conversational agent to enhance second language learners’ willingness to communicate,'' \emph{International Journal of Artificial Intelligence in Education}, vol.~29, no.~1, pp. 29--57, 2019.

\bibitem{lee2002eyes}
S.~P. Lee, J.~B. Badler, and N.~I. Badler, ``Eyes alive,'' in \emph{Proceedings of the 29th annual conference on Computer graphics and interactive techniques}, 2002, pp. 637--644.

\bibitem{yalccin2020empathy}
{\"O}.~N. Yal{\c{c}}{\i}n, ``Empathy framework for embodied conversational agents,'' \emph{Cognitive Systems Research}, vol.~59, pp. 123--132, 2020.

\bibitem{sajjadi2019personality}
P.~Sajjadi, L.~Hoffmann, P.~Cimiano, and S.~Kopp, ``A personality-based emotional model for embodied conversational agents: Effects on perceived social presence and game experience of users,'' \emph{Entertainment Computing}, vol.~32, p. 100313, 2019.

\bibitem{martinez2020multiparty}
V.~R. Martinez and J.~Kennedy, ``A multiparty chat-based dialogue system with concurrent conversation tracking and memory,'' in \emph{Proceedings of the 2nd Conference on Conversational User Interfaces}, 2020, pp. 1--9.

\bibitem{andreotti2021perception}
L.~Andreotti, M.~L. Weber, T.~L. da~Silva, V.~F. de~Andrade~Araujo, and S.~R. Musse, ``Perception of charisma, comfort, micro and macro expressions in computer graphics characters,'' in \emph{2021 20th Brazilian Symposium on Computer Games and Digital Entertainment (SBGames)}.\hskip 1em plus 0.5em minus 0.4em\relax IEEE, 2021, pp. 107--116.

\bibitem{zell2019perception}
E.~Zell, K.~Zibrek, and R.~McDonnell, ``Perception of virtual characters,'' in \emph{ACM SIGGRAPH 2019 Courses}, 2019, pp. 1--17.

\bibitem{katsyri2015review}
J.~K{\"a}tsyri, K.~F{\"o}rger, M.~M{\"a}k{\"a}r{\"a}inen, and T.~Takala, ``A review of empirical evidence on different uncanny valley hypotheses: support for perceptual mismatch as one road to the valley of eeriness,'' \emph{Frontiers in psychology}, vol.~6, p. 390, 2015.

\bibitem{shahid2014no}
M.~Shahid, A.~Rossholm, B.~L{\"o}vstr{\"o}m, and H.-J. Zepernick, ``No-reference image and video quality assessment: a classification and review of recent approaches,'' \emph{EURASIP Journal on image and Video Processing}, vol. 2014, no.~1, p.~40, 2014.

\bibitem{tinwell2011facial}
A.~Tinwell, M.~Grimshaw, D.~A. Nabi, and A.~Williams, ``Facial expression of emotion and perception of the uncanny valley in virtual characters,'' \emph{Computers in Human Behavior}, vol.~27, no.~2, pp. 741--749, 2011.

\bibitem{bailenson2005independent}
J.~N. Bailenson, K.~Swinth, C.~Hoyt, S.~Persky, A.~Dimov, and J.~Blascovich, ``The independent and interactive effects of embodied-agent appearance and behavior on self-report, cognitive, and behavioral markers of copresence in immersive virtual environments,'' \emph{Presence: Teleoperators \& Virtual Environments}, vol.~14, no.~4, pp. 379--393, 2005.

\bibitem{gouskos2006depths}
C.~Gouskos, ``The depths of the uncanny valley,'' \emph{DOI= http://uk. gamespot. com/features/6153667/index. html}, 2006.

\bibitem{dill2012evaluation}
V.~Dill, L.~M. Flach, R.~Hocevar, C.~Lykawka, S.~R. Musse, and M.~S. Pinho, ``Evaluation of the uncanny valley in cg characters,'' in \emph{International Conference on Intelligent Virtual Agents}.\hskip 1em plus 0.5em minus 0.4em\relax Springer, 2012, pp. 511--513.

\bibitem{beghdadi2013survey}
A.~Beghdadi, M.-C. Larabi, A.~Bouzerdoum, and K.~M. Iftekharuddin, ``A survey of perceptual image processing methods,'' \emph{Signal Processing: Image Communication}, vol.~28, no.~8, pp. 811--831, 2013.

\bibitem{Theodoridis2014}
S.~Theodoridis and R.~Chellappa, \emph{Image and Video Compression and Multimedia}.\hskip 1em plus 0.5em minus 0.4em\relax Academic Press, 2014.

\bibitem{dal2021can}
G.~P. Dal~Molin, F.~M. Nomura, B.~M. Dalmoro, V.~F. d.~A. Ara{\'u}jo, and S.~R. Musse, ``Can we estimate the perceived comfort of virtual human faces using visual cues?'' in \emph{2021 IEEE 15th International Conference on Semantic Computing (ICSC)}.\hskip 1em plus 0.5em minus 0.4em\relax IEEE, 2021, pp. 366--369.

\bibitem{tumblin2001applied}
J.~Tumblin and J.~A. Ferwerda, ``Applied perception,'' \emph{IEEE Computer Graphics and Applications}, vol.~21, no.~5, pp. 20--21, 2001.

\bibitem{dal2022estimating}
G.~P. Dal~Molin, V.~F. de~Andrade~Araujo, and S.~R. Musse, ``Estimating perceived comfort in virtual humans based on spatial and spectral entropy.'' in \emph{VISIGRAPP (4: VISAPP)}, 2022, pp. 436--443.

\bibitem{liu2014no}
L.~Liu, B.~Liu, H.~Huang, and A.~C. Bovik, ``No-reference image quality assessment based on spatial and spectral entropies,'' \emph{Signal Processing: Image Communication}, vol.~29, no.~8, pp. 856--863, 2014.

\bibitem{laue2017familiar}
C.~Laue, ``Familiar and strange: Gender, sex, and love in the uncanny valley,'' \emph{Multimodal technologies and interaction}, vol.~1, no.~1, p.~2, 2017.

\bibitem{kteily2022dehumanization}
N.~S. Kteily and A.~P. Landry, ``Dehumanization: trends, insights, and challenges,'' \emph{Trends in cognitive sciences}, 2022.

\bibitem{araujo2022towards}
V.~Araujo, D.~Schaffer, A.~B. Costa, and S.~R. Musse, ``Towards virtual humans without gender stereotyped visual features,'' in \emph{SIGGRAPH Asia 2022 Technical Communications}, 2022, pp. 1--4.

\bibitem{mcdonnell2007virtual}
R.~McDonnell, S.~J{\"o}rg, J.~K. Hodgins, F.~Newell, and C.~O'Sullivan, ``Virtual shapers \& movers: form and motion affect sex perception,'' in \emph{Proceedings of the 4th symposium on Applied perception in graphics and visualization}, 2007, pp. 7--10.

\bibitem{araujo2021analysis}
V.~Araujo, B.~Dalmoro, and S.~R. Musse, ``Analysis of charisma, comfort and realism in cg characters from a gender perspective,'' \emph{The Visual Computer}, pp. 1--14, 2021.

\bibitem{kim2022countering}
T.~Kim, H.~Rushmeier, J.~Dorsey, D.~Nowrouzezahrai, R.~Syed, W.~Jarosz, and A.~Darke, ``Countering racial bias in computer graphics research,'' in \emph{ACM SIGGRAPH 2022 Talks}, 2022, pp. 1--2.

\bibitem{condry1976sex}
J.~Condry and S.~Condry, ``Sex differences: A study of the eye of the beholder,'' \emph{Child development}, pp. 812--819, 1976.

\bibitem{helbing1995social}
D.~Helbing and P.~Molnar, ``Social force model for pedestrian dynamics,'' \emph{Physical review E}, vol.~51, no.~5, p. 4282, 1995.

\bibitem{van2011reciprocal}
J.~Van Den~Berg, S.~J. Guy, M.~Lin, and D.~Manocha, ``Reciprocal n-body collision avoidance,'' in \emph{Robotics Research: The 14th International Symposium ISRR}.\hskip 1em plus 0.5em minus 0.4em\relax Springer, 2011, pp. 3--19.

\bibitem{de2012simulating}
A.~de~Lima~Bicho, R.~A. Rodrigues, S.~R. Musse, C.~R. Jung, M.~Paravisi, and L.~P. Magalh{\~a}es, ``Simulating crowds based on a space colonization algorithm,'' \emph{Computers \& Graphics}, vol.~36, no.~2, pp. 70--79, 2012.

\bibitem{pelechano2005crowd}
N.~Pelechano~G{\'o}mez, K.~O'Brien, B.~G. Silverman, and N.~Badler, ``Crowd simulation incorporating agent psychological models, roles and communication,'' in \emph{First International Workshop on Crowd Simulation}, 2005.

\bibitem{durupinar2015psychological}
F.~Durup{\i}nar, U.~G{\"u}d{\"u}kbay, A.~Aman, and N.~I. Badler, ``Psychological parameters for crowd simulation: From audiences to mobs,'' \emph{IEEE transactions on visualization and computer graphics}, vol.~22, no.~9, pp. 2145--2159, 2015.

\bibitem{xiong2010hybrid}
M.~Xiong, M.~Lees, W.~Cai, S.~Zhou, and M.~Y.~H. Low, ``Hybrid modelling of crowd simulation,'' \emph{Procedia Computer Science}, vol.~1, no.~1, pp. 57--65, 2010.

\bibitem{da2020towards}
A.~da~Silva~Antonitsch, S.~R. Musse, and L.~H. de~Figueiredo, ``Towards a legion of virtual humans: Steering behaviors and organic visualization,'' in \emph{2020 33rd SIBGRAPI Conference on Graphics, Patterns and Images (SIBGRAPI)}.\hskip 1em plus 0.5em minus 0.4em\relax IEEE, 2020, pp. 31--38.

\bibitem{radford2018improving}
A.~Radford, K.~Narasimhan, T.~Salimans, I.~Sutskever \emph{et~al.}, ``Improving language understanding by generative pre-training,'' 2018.

\bibitem{biancardi2019computational}
B.~Biancardi, C.~Wang, M.~Mancini, A.~Cafaro, G.~Chanel, and C.~Pelachaud, ``A computational model for managing impressions of an embodied conversational agent in real-time,'' in \emph{2019 8th International Conference on Affective Computing and Intelligent Interaction (ACII)}.\hskip 1em plus 0.5em minus 0.4em\relax IEEE, 2019, pp. 1--7.

\bibitem{kramer2019developing}
L.~Kramer, S.~ter Stal, B.~Mulder, E.~de~Vet, and L.~van Velsen, ``Developing embodied conversational agents for healthy lifestyles: a scoping review,'' in \emph{ARPH}, 2019.

\bibitem{philip2020trust}
P.~Philip, L.~Dupuy, M.~Auriacombe, F.~Serre, E.~de~Sevin, A.~Sauteraud, and J.-A. Micoulaud-Franchi, ``Trust and acceptance of a virtual psychiatric interview between embodied conversational agents and outpatients,'' \emph{npj Digital Medicine}, vol.~3, no.~1, pp. 1--7, 2020.

\bibitem{knob2023arthur}
P.~R. Knob, N.~D. Pizzol, S.~R. Musse, and C.~Pelachaud, ``Arthur and bella: multi-purpose empathetic ai assistants for daily conversations,'' \emph{The Visual Computer}, pp. 1--16, 2023.

\end{thebibliography}

\end{document}